\documentclass[twocolumn]{aastex62}

\graphicspath{{./}{figures/}}
\usepackage{graphicx}
\usepackage{subfigure}
\usepackage{placeins}
\usepackage{lineno}
\usepackage{hhline}
\usepackage{media9}

\usepackage{booktabs}
\usepackage{amsmath}

\usepackage{commath}

\usepackage{braket}

\usepackage{xcolor}
\received{}
\revised{}
\accepted{}

\submitjournal{ApJ}

\shorttitle{Sample article}
\shortauthors{Tajfirouze et al.}

\begin{document}

\title{Spatially Coherent and Intermittent Alfv\'enic Fluctuations in  Solar Polar Spicules}

\correspondingauthor{Edris Tajfirouze}
\email{e.tajfirouzeh@qmul.ac.uk }

\author[0000-0002-0786-7307]{Edris Tajfirouze}
\affil{Department of Physics and Astronomy, Queen Mary University of London, London, E1 4NS, UK}

\author{Christopher H. K. Chen}
\affiliation{Department of Physics and Astronomy, Queen Mary University of London, London, E1 4NS, UK}

\author{Richard J. Morton}
\affiliation{Department of Mathematics, Physics and Electrical Engineering, Northumbria University, Newcastle upon Tyne, UK.}

\author{Peter R. Young}
\affiliation{Department of Mathematics, Physics and Electrical Engineering, Northumbria University, Newcastle upon Tyne, UK.}
\affiliation{Heliophysics Division, NASA Goddard Space Flight Center, Greenbelt, MD, 20771, USA.}

\begin{abstract}
Alfv\'enic fluctuations are considered a key mechanism for transporting energy from the lower solar atmosphere into the corona, with spicules acting as dynamic conduits for this transfer. We investigate transverse and Doppler velocity fluctuations in quiet-Sun polar spicules observed in the Si IV 1394~\AA\ line by the Interface Region Imaging Spectrograph. Fourier analysis in both time and space is used to characterize the distribution of power across frequency and spatial scales. The temporal power spectra show broadband fluctuations, with a broad power enhancement in the 3–7 mHz range and maximum power near 4–6 mHz. Spatial Fourier analysis of Doppler velocity fluctuations reveals broadband behavior with a perpendicular power spectrum scaling as $\sim k_\perp^{-1.43}$, similar to but slightly shallower than the canonical spectral index values of $-5/3$ and $-3/2$ expected for homogeneous strong MHD turbulence, but within the range of those found in simulations of reflection-driven turbulence in the solar corona. Velocity increment probability distribution functions exhibit non-Gaussian behavior, with the kurtosis near Gaussian at large spatial scales and increasing toward smaller scales following a power-law scaling (index $\approx -0.23$), consistent with the development of intermittency. Spatial coherence analysis based on cross-correlation, together with spectral and kurtosis-based diagnostics, indicates a common outer-scale range of about a few hundred to about a thousand kilometres, with the cross-correlation method yielding smaller estimates. These results provide new observational evidence that, in addition to coherent Alfv\'en waves,
polar spicules host a multiscale Alfv\'enic fluctuations with characteristics consistent with a developing cascade, MHD turbulence, and intermittency, highlighting their potential role in mediating energy transport into the solar corona.

\end{abstract}


\section{Introduction} \label{sec:intro}

The solar corona, with temperatures exceeding a million degrees, is sustained by energy originating in the photosphere and mediated through the chromosphere. However, despite considerable observational and modeling efforts, the mechanisms by which the dynamics and energetics of the chromosphere and corona are coupled remain poorly understood \citep{Klimchuk_2006, Hansteen_etal_2007, Samanta_etal_2019}.

Spicules have emerged as a leading candidate in facilitating the exchange of mass and energy between the lower solar atmosphere and the corona \citep{Beckers_1968, PK_1978, Athay_Holzer_1982, DePontieu_etal_2009}. These highly dynamic, jet-like structures, channel plasma and momentum upward often reaching coronal heights where they appear as elongated features at the limb, providing natural pathways for mass and energy transport, including wave energy, and linking small scale chromospheric dynamics to large scale coronal energetics.
% Although, there are several observations show that spicules can reach coronal temperature and deposit thermal energy directly in the corona.

In particular, Alfv\'enic\footnote { The term  Alfv\'enic describes magnetohydrodynamic waves in inhomogeneous plasma that
retain the essential properties of pure Alfv\'en waves. These modes are the generalisation of surface Alfv\'en waves \citep{Wentzel_1979, Roberts_1981} and  encompass the MHD kink mode in structured plasma \citep{Goossens_etal_2012}. } waves traveling along spicules have received significant attention because they can carry substantial Poynting flux over extended distances. While partially reflected and dissipated during their propagation, these waves are capable of transferring energy from the convective motions in the photosphere out to the corona and deliver energy that may contribute significantly to coronal heating and solar wind acceleration \citep{McIntosh_etal._2011, VanD_etal_2020}.

Although the exact energy flux carried by these waves remains debated, estimates suggest it is comparable to that required to sustain the quiet-Sun corona and drive the solar wind, approximately 200 W m$^{-2}$ \citep{WN_1977, Aschwanden_etal_2007, DePontieu_etal._2007}.

 \begin{figure*}[tp]
  \epsscale{.9}
   \plotone{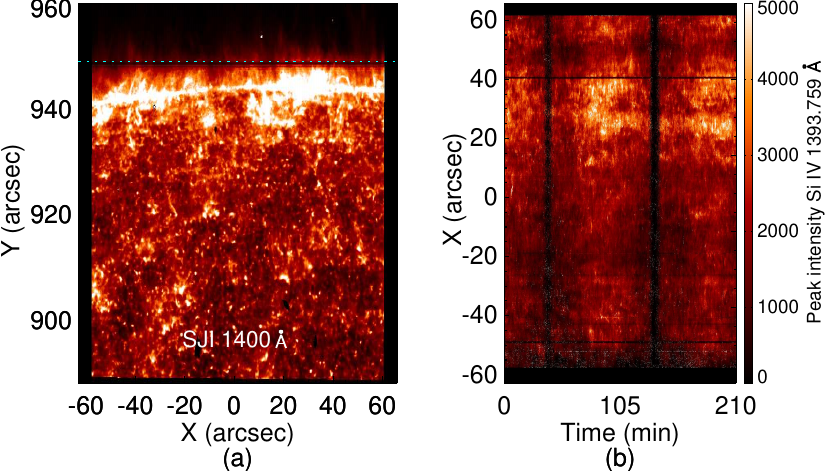}
     \caption{Panel (a) displays a raster scan of the region from IRIS at Si IV $\lambda$1400~\AA. The green dotted line indicates the location of the sit-and-stare observations. Panel (b)  shows the peak line intensity for Si IV $\lambda$~1393.75~{\AA} from the sit-and-stare observations.}
   \label{fig:fig1}
 \end{figure*}
 
Propagating Alfv\'enic fluctuations have been observed for over a decade, both in the low corona \citep[e.g.,][]{Tomczyk_etal._2007, McIntosh_etal._2011, Morton_etal_2015, Morton_etal_2025, Tajfirouze_etal_2025} and in chromospheric spicules \citep{He_etal_2009, Okamoto_DePontieu_2011, Morton_etal_2013}. These fluctuations typically exhibit periods of 3–5 minutes, coinciding with the dominant timescales of photospheric motions, and can be most effectively analyzed through their power spectra.  Measurements reveal broadband power-law distributions across frequencies, often with pronounced enhancements near 3–5 mHz, with slopes varying depending on local magnetic topology \citep{Morton_etal_2015}. Previous studies propose that this wave energy originates from double mode conversion of internal acoustic p-modes \citep{Cally_Goossens_2008}, while more recent work indicates that filtering of photospheric wave power by frequency-dependent transmission in the lower atmosphere dominates, with p-mode conversion playing only a secondary role \citep{Morton_Soler_2025, Soler_etal_2019}.

 While temporal power spectra characterize how wave energy is distributed across frequencies, they do not constrain how that energy is distributed across physical scales. In a structured and magnetized plasma, such as the solar corona, the spatial distribution of velocity fluctuations contains critical information about wave evolution, energy transfer, and dissipation. If Alfv\'enic fluctuations undergo nonlinear interactions, their energy is expected to cascade toward smaller spatial scales, producing scale-dependent structure that can be quantified through spatial Fourier analysis (\citealt{Iroshnikov_1964}, \citealt{Kraichnan_1965}, \citealt{GS_1995}, \citealt{Velli_1993}, and \citealt{VanB_2011}; see also the review by \citealt{Schekochihin_2022}). In magnetohydrodynamic systems, turbulent cascades are commonly associated with power-law spatial spectra, reflecting the scale-invariant transfer of energy across spatial scales. Such behavior has been widely observed in the solar wind \citep[see, e.g., ][]{BC_2013, Chen_2016, Verscharen_etal_2019} and is increasingly suggested in coronal observations \citep{Morton_etal._2016, Battams_etal_2017}. Identifying these scaling behaviors is essential to determine whether observed Alfv\'enic motions act merely as propagating waves or participate in a broader turbulent energy transfer process.

Beyond scale-dependent power, turbulence is fundamentally associated with intermittency, the uneven spatial distribution of fluctuations and the presence of localized, intense gradients \citep{Frisch_1995}. Intermittent systems depart from Gaussian statistics, particularly at small spatial separations where coherent structures and nonlinear interactions dominate. By combining spatial Fourier analysis with higher-order statistics such as velocity increment probability distribution functions and kurtosis, we can assess whether observed Alfv\'enic motions participate in a turbulent cascade, providing a unified view of the frequency distribution, spatial organization, and statistical structure of fluctuations in the outer solar atmosphere \citep{Sorriso-Valvo_etal_2001, salem_etal_2009, osman_etal_2011, Chen_etal_2014}.

In addition to spectral analysis, the spatial coherence of transverse motions provides critical insight into how wave energy is organized and transported from spicules into coronal structures. Observations further reveal that Alfv\'enic oscillations in the corona can remain coherent over scales larger than the width of an individual loop, indicating that neighboring loops can oscillate collectively as bundles \citep{ShM_2023, Tajfirouze_etal_2025, Hahn_etal_2025, Hahn2_etal_2025}. The correlation length scale quantifies the characteristic spatial scale of energy input, providing an essential observational constraint for modeling of Alfv\'en wave turbulence, a leading candidate for the dissipation of Alfv\'enic waves. In theory, such turbulence can be driven either by nonlinear interaction of counterpropagating waves in a longitudinally stratified atmosphere \citep{Velli_1993, CV_2005, Matthaeus_etal_1999, Matthaeus_etal_2003, Dmitruk_etal_2001, Perez_Chandran_2013, vanderHolst_etal_2014, CH-P_2019, Chandran_etal_2025, Tajfirouze_etal_2025}, or by nonlinear deformation of the wave packets in a transversely structured plasma \citep{Magyar_etal_2019}. 
 
Here we analyze transverse oscillations in quiet-Sun polar spicules using IRIS observations \citep{DePontieu_etal_2014}, whose design specifically targets the dynamic interface between the chromosphere and corona, making it ideally suited for spicule studies.
IRIS provides diagnostics of the chromosphere and transition region through a number of strong UV lines, achieving relative Doppler shift measurements accurate to better than 1 km s$^{-1}$ \citep{DePontieu_etal_2014}.
The high spatial resolution ($0.33\arcsec$–$0.4\arcsec$) and flexible temporal sampling (with cadences shorter than 2 s achievable in some observing modes) of IRIS make it possible to resolve fine-scale dynamics in the chromosphere and transition region, allowing the detection of oscillations and waves with unprecedented clarity. For the observations analyzed here, the cadence is approximately 9.4 s. In this work, we perform a combined temporal and spatial analysis of Alfv\'enic fluctuations in spicules. Using Fourier and higher-order statistical diagnostics, we characterize their spectral properties, transverse coherence, and intermittency; this provides new observational evidence for intermittent, turbulence-like energy transfer in these structures and offers a more complete description of their multi-scale behavior than has been achieved in previous studies.

\medskip

\section{Observation and Data Reduction}

The Interface Region Imaging Spectrograph \citep[IRIS;][]{DePontieu_etal_2014} performs coordinated ultraviolet spectroscopic and imaging observations of the solar atmosphere. A slit-jaw imager (SJI) provides simultaneous context imaging over a $175\arcsec \times 175\arcsec$ field of view in four passbands centered on C II 1335 \AA, Si IV 1400 \AA, Mg II k 2796 \AA, and the 2830 \AA\ continuum.
Its spectrograph uses a narrow slit ($0.33\arcsec–0.4\arcsec$ wide and $75\arcsec$ long) that records spectra in two 
far-ultraviolet (FUV; 1332–1358 \AA\ and 1389–
1407 \AA) and one near-ultraviolet (NUV; 2783–2835 \AA) wavelength bands, covering emission lines formed from the photosphere ($\sim$5 $\times$ 10$^{3}$ K) to the low corona ($\sim$10$^{6}$ K). In spectroscopic mode, IRIS can operate in raster or sit-and-stare modes, where the slit is either scanned across the target region or kept fixed while spacecraft tracking maintains a stable pointing.

In the present work, we focus on sit-and-stare observations of the quiescent corona at the solar north polar limb. The data analyzed were obtained on 17 June 2014, starting at 07:29:45 UT. The observations were made using a slit oriented parallel to the limb in the east-west direction, positioned off-limb to the north.
The slit covers a region of $0.33\arcsec$ $\times$ $175\arcsec$ with a cadence $\sim$ 9.4 s. The location of the slit center is X=$-2.64\arcsec$ and Y=$952.75\arcsec$.
% confirming that the observational field of view is  located in the corona (slit center located at $\sim5000$~km above the photosphere). 

\begin{figure*}[tp]
 \epsscale{.9}
\plotone{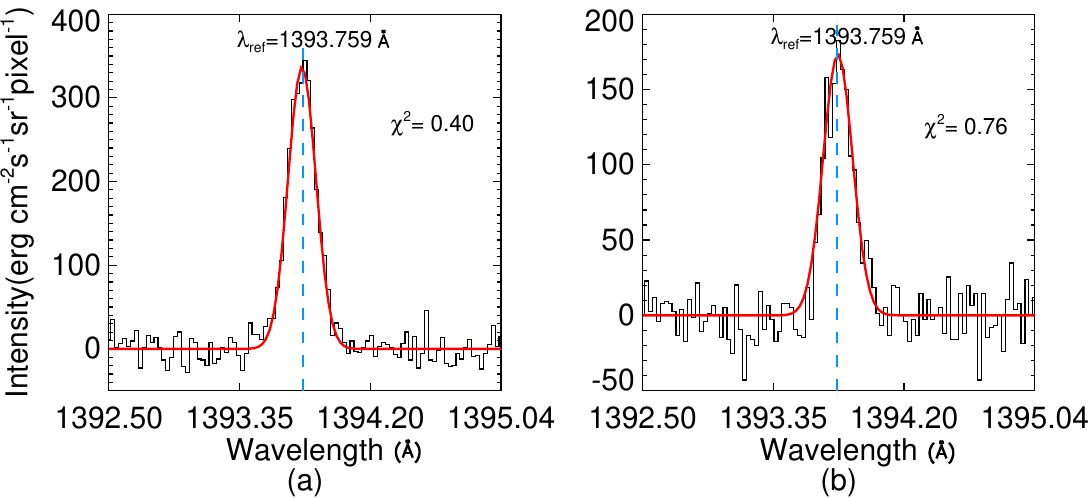}
    \caption{Two examples of spectra with Gaussian fits over-plotted with a red solid line. Panel (a) shows the spectrum at position $x = 30.21\arcsec$ and time $t = 1411.7\,\mathrm{s}$ with a Gaussian fit yielding $\chi^2=0.4$. Panel (b) shows the spectrum at position $x = -50.3\arcsec$ and time $t = 2277.3\,\mathrm{s}$ with a Gaussian fit yielding $\chi^2=6.83$. The dashed blue line shows the reference wavelength.}
  \label{fig:fig_spectra}
\end{figure*}

Figure \ref{fig:fig1} shows the slit position of IRIS on an IRIS slit-jaw image. 

Assuming that the bright ring we see in 1400 \AA\ is just above the photospheric limb within $1\arcsec$ and thinking of a stratified atmosphere \citep{VAL_1981}, where the chromosphere only reaches up to 2 Mm heights and above is the transition region (for about 100 km) and then the corona, it is evident that the slit is sampling the tops of the spicule forest, so within the coronal volume, but plasma at cooler temperatures. 
We notice that small-scale, low-lying transition-region loops have previously been reported near the solar limb \citep{Hansteen_etal_2014}, and such structures are also visible in Figure~\ref{fig:fig1}a. The slit position was selected to focus primarily on regions extending above the limb where spicule-like structures are expected to dominate, although some contribution from low-lying loop structures along the line of sight cannot be entirely excluded.

In projected geometry, the midpoint of the slit is located at an altitude of approximately $\sim$ 4500 km above the photosphere. 

The magnetic field lines associated with the spicules are predominantly radial at the slit height, implying that the slit is approximately oriented perpendicular to the projected magnetic field direction in the plane of the sky. Under this viewing geometry, line-of-sight Doppler velocity variations are interpreted as signatures of transverse displacements relative to the magnetic field, while spatial variations along the slit are interpreted as structure perpendicular to the field-aligned direction.
We select Si IV $\lambda$~1393.75~\AA\ data to analyze as it is one of the strongest lines in the IRIS FUV spectra. This line is typically optically thin, and provides robust diagnostics of transition region structuring and dynamics. 

\begin{figure*}[tp]
 \epsscale{0.9}
\plotone{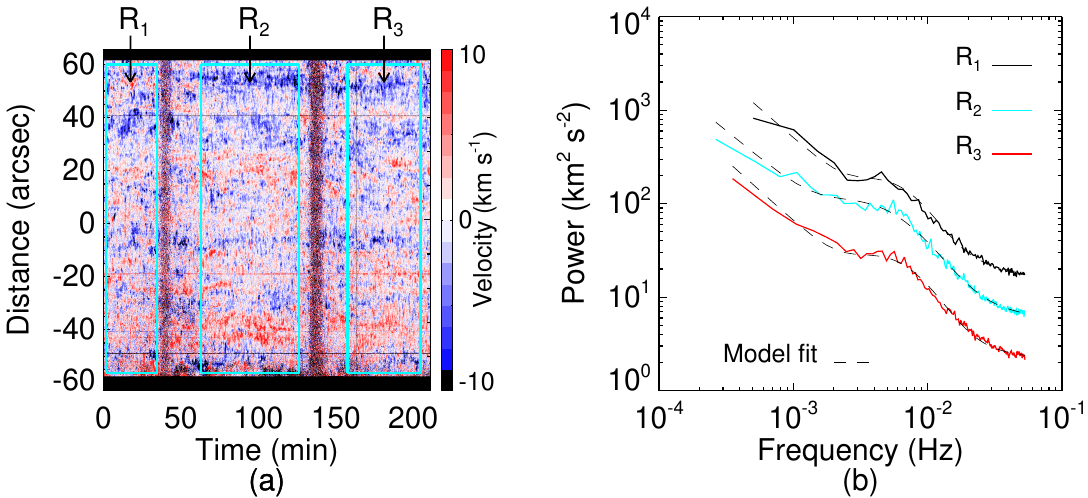}
    \caption{Panel (a) shows the Doppler velocity map derived from the emission line fit, with three distinct boxes indicating the regions where power spectra were calculated. The Doppler velocity map has been resampled and interpolated to fill missing values. Panel (b) displays the power spectra of the Doppler velocities from the selected regions, with the fitted models over-plotted as black dashed lines. To avoid overlap and improve clarity, the spectra from the second and third regions have been multiplied by factors of 0.6 and 0.2, respectively, while the first spectrum is shown without scaling. The vertical dotted line marks 4.5 mHz, near the frequency of maximum power within the broad spectral enhancement.}
  \label{fig:fig_Doppler}
\end{figure*}

 We analyze level 2 IRIS data which are pre-processed to correct for slit tilt and orbit variations. The Si IV $\lambda$~1393.75~\AA\ emission line is fitted with a single Gaussian profile.
 The line profiles in the present dataset are generally well described by this approach, with only a small fraction exhibiting significant deviations from Gaussianity. Following \citep{Rao_etal_2022, KP_2023}, we adopt a reduced $\chi^2$ threshold of 5 to identify poor Gaussian representations of the observed profiles. We find that only about 5\% of the fitted profiles exceed this threshold, indicating that the vast majority of the dataset is consistent with near-Gaussian line shapes. This suggests that strongly non-Gaussian features, such as pronounced asymmetries or multi-component structures, are relatively rare in the observations. For the bulk of the data, the Gaussian-derived centroid provides a reliable estimate of the line-of-sight Doppler shift, while caution is required only for the small subset of profiles with elevated $\chi^2$ values. To illustrate this behavior, we show example spectra in Figure~\ref{fig:fig_spectra} extracted at different positions along the slit, highlighting cases with both low and high $\chi^2$ values and demonstrating the corresponding variation in profile quality. We further verified that masking these outlier pixels and interpolating over them does not significantly affect the derived results, indicating that the main conclusions are robust to the treatment of these non-Gaussian profiles.
 
 From the fit parameters we extracted the Doppler shifts of the line center.
This line is typically optically thin, implying that the observed emission represents a superposition of structures along the line of sight. Consequently, even regions with relatively weak Si IV intensity are expected to contain line-of-sight integrated emission from unresolved fine structures associated with the spicule forest, rather than representing emission-free plasma volumes.
 Our analysis primarily targets variations in Doppler shifts; therefore, precise determination of the absolute reference wavelengths is unnecessary. IRIS observations are stabilized by an onboard image stabilization system, and therefore no additional jitter correction is required. 

\medskip

\section{Results}

% \begin{table*}[t!]
%     \centering
%     \begin{adjustbox}{max width=\textwidth}
%     \begin{tabular}{ c|c|c|c|c|c|c }
%        \hline
%         \hline
%        Model &  
%        \multicolumn{6}{c}{Fit parameters \& Standard errors $\sigma$}\\
%          & \ \ $A$ \ \ , \ \ $\sigma_{A}$  \ \ & \ \ $\alpha$ \ \ , \ \ $ \sigma_{\alpha} $ \ \ & \ \ $B$\ \ , \ \ $\sigma_{B}$ \ \  & \ \ $C$ \ \ , \ \ $\sigma_{C}$ \ \  & \ \ $D$ \ \ , \ \ $\sigma_{D}$ \ \ & \ \ $E$ \ \ , \ \ $\sigma_{E}\ \ $ \\
            
%         \hline
%          Model fit in R$_{1}$   &  0.07 , 0.2 & -1.27 , 0.39 & 15.02 , 5.86 & 95.32 , 39.93 & -2.29 , 0.13 & 0.23 , 0.03 \\
%         Model fit in R$_{2}$  & 0.07 , 0.1 & -1.17 , 0.16 & 8.73 , 1.72 & 102.41 , 32.27 & -2.43 , 0.16& -0.31 , 0.07\\
       
%         Model fit in R$_{3}$  & 0.03 , 0.03 & -1.3 , 0.14 & 10.12 , 1.76 & 82.45 , 31.83 & -2.3 , 0.13& 0.25 , 0.07\\
%          % &  &  &  & & & \\
        
%         \hline
%          \hline
%     \end{tabular}
%     \end{adjustbox}
%     \caption{Model fit parameters}
%     \label{tab:fitparam}
% \end{table*}

\begin{table*}[t!]
    \centering

    \resizebox{\textwidth}{!}{%
    \begin{tabular}{c|c|c|c|c|c|c}
       \hline
       \hline
       Model &  
       \multicolumn{6}{c}{Fit parameters \& Standard errors $\sigma$}\\
         & $A , \sigma_{A}$ & $\alpha , \sigma_{\alpha}$ & $B , \sigma_{B}$ & $C , \sigma_{C}$ & $D , \sigma_{D}$ & $E , \sigma_{E}$ \\
        \hline
        Model fit in R$_{1}$ & 0.07 , 0.2 & -1.27 , 0.39 & 15.02 , 5.86 & 95.32 , 39.93 & -2.29 , 0.13 & 0.23 , 0.03 \\
        Model fit in R$_{2}$ & 0.07 , 0.1 & -1.17 , 0.16 & 8.73 , 1.72 & 102.41 , 32.27 & -2.43 , 0.16 & -0.31 , 0.07\\
        Model fit in R$_{3}$ & 0.03 , 0.03 & -1.3 , 0.14 & 10.12 , 1.76 & 82.45 , 31.83 & -2.3 , 0.13 & 0.25 , 0.07\\
        \hline
        \hline
    \end{tabular}%
    }

    \caption{Model fit parameters}
    \label{tab:fitparam}
\end{table*}

From the Gaussian fit parameters of the spectral line, we derived the Doppler velocity. Figure~\ref{fig:fig_Doppler} shows the Doppler velocity time-distance diagram. The original Doppler velocity maps contain two columns of missing pixels, which we attribute to particle hits associated with passages through the South Atlantic Anomaly (SAA). To avoid introducing artifacts into the analysis, these regions were excluded, and three clean subfields (indicated by the green rectangular boxes were selected for further study.). Even within the selected areas, occasional missing or corrupted pixels remain. 
To mitigate the impact of occasional corrupted pixels and failed Gaussian fits, we applied an amplitude threshold to the Doppler velocities ($|v|<20$ km s$^{-1}$). Values exceeding this limit were replaced using linear interpolation. The threshold was chosen based on inspection of the velocity distribution, for which values beyond the adopted limit constitute a relatively small population compared with the bulk of the distribution that is consistent with outliers associated with poor spectral fits or low-signal measurements rather than the bulk of the observed fluctuations. This procedure preserves the underlying temporal variability while ensuring that the data are suitable for Fourier and time-series analysis.We note that the choice of threshold can influence higher-order statistics. If the threshold is set too high, isolated outliers can artificially enhance the tails of the velocity-increment distributions and inflate kurtosis estimates, whereas an overly restrictive threshold may remove genuine large-amplitude fluctuations and reduce the inferred intermittency. We therefore tested a range of threshold values and found that the temporal and spatial power spectra, correlation-length estimates, and the scale-dependent increase of kurtosis toward small spatial scales remain qualitatively unchanged. The adopted threshold may affect the absolute value of kurtosis at some scales, but does not alter the main conclusions regarding the presence of non-Gaussian statistics and intermittency.

% To mitigate their impact, we applied an amplitude threshold to the Doppler velocities ($|v| < $20 km s$^{-1}$). Values exceeding this limit were filled using linear interpolation. This procedure preserves the underlying temporal variability while ensuring that the data are suitable for Fourier and time–series analysis. We note that this threshold suppresses rare, high-amplitude fluctuations and may lead to a modest underestimation of intermittency diagnostics (e.g., kurtosis), while having a limited impact on the dominant coherent fluctuations considered here.

We also note a systematic decrease in intensity along the slit in the few minutes before and after SAA passages. This is likely caused by a small northward pointing drift of IRIS during these intervals, as can be seen in the slit-jaw image sequences. Since the structures of interest are assumed to be predominantly radial, this drift is not expected to introduce large artificial Doppler signatures. Nevertheless, a displacement of the slit relative to the limb may alter the plasma population sampled during these intervals and could therefore affect the measured Doppler statistics, spatial correlations, and power spectra. We regard this as a limitation of the observations. However, the close agreement of the spectral slopes, coherence estimates, and intermittency diagnostics obtained from the three independently analyzed regions suggests that any such effect is not a dominant contributor to the results presented here.

The root-mean-square (RMS) amplitudes of the Doppler velocity fluctuations measured in our IRIS observations range from 2–8~km s$^{-1}$, somewhat larger than the typical values of 0.5–2~km s$^{-1}$ commonly reported for Alfv\'enic wave signatures in spectroscopic data \citep{Tomczyk_etal._2007, Tian_etal_2012, Morton_etal_2025, Tajfirouze_etal_2025}. Nevertheless, they remain smaller than the 15–30~km s$^{-1}$ transverse amplitudes inferred from imaging observations \citep[e.g.,][]{McIntosh_etal._2011, Thurgood_etal_2014, Morton_etal._2019}. We note, however, that IRIS spectroscopic observations of spicules have also revealed Alfv\'enic motions with amplitudes approaching $\sim$20 km s$^{-1}$, inferred from substantial non-thermal line broadening \citep{Tian_etal_2014}. As discussed extensively in previous work, this discrepancy is expected because Doppler measurements sample an optically thin plasma and therefore suffer from line-of-sight integration, which averages over multiple out-of-phase wave motions and naturally reduces the observed velocity amplitudes \citep[see, e.g., ][]{DeM_Pasco_2012, McI_DeP_2012, DP_2012, Morton_etal._2016, Pant_etal_2019}. 

The data exhibit a reasonably regular cadence, with variations of up to 0.13 seconds between exposures. To standardize the time sampling, we resample the velocity time series to a uniform cadence equal to the average cadence, using linear interpolation. This resampling process effectively smooths the data locally in time, thereby reducing high-frequency noise.

\subsection{Temporal Analysis}
\subsubsection{Frequency Power Spectrum}

We perform Fourier analysis on the Doppler velocity time series derived from the Gaussian fits.
Subtracting either the temporal or spatial mean produces no appreciable change in the resulting Fast Fourier Transform (FFT) or power spectra, indicating that the velocity signals do not contain significant large-scale systematic trends that would affect their spectral properties.

To minimize spectral leakage, each time series was multiplied by a Hanning window before computing the FFT. Power spectra were computed at each spatial location along the slit and then averaged over the entire slit for each analyzed time interval to obtain representative spectra.

The resulting spectra (Figure~\ref{fig:fig_Doppler}(b)) exhibit a broadband power-law distribution with broad enhancement in the frequency range of approximately 3–7 mHz, with maximum power occurring near 4–6 mHz. This behavior is consistent with previous spectroscopic measurements of Alfv\'enic velocity fluctuations in the solar corona, which similarly show power-law scaling with enhanced power near photospheric oscillation frequencies.

The power spectra derived from the analyzed intervals are nearly identical in shape and follow the same power-law behavior. This consistency indicates that the system remains in a quasi-stationary regime, with the observed velocity fluctuations governed by the same underlying physical process throughout the observation.

We note that some of the low-frequency fluctuations correspond to periods comparable to, or longer than, typical Type-II spicule lifetimes. Therefore, these signals should not be interpreted as coherent oscillations of individual spicules persisting over multiple wave cycles. Instead, because the IRIS slit samples a continuously evolving ensemble of spicular plasma, the temporal power spectrum represents the statistical properties of velocity fluctuations within the observed spicule population. Long-period power may therefore reflect persistent or recurrent driving acting across successive spicules rather than oscillations of a single structure. Similar lifetime-related considerations have been discussed in previous studies of waves in spicules \citep[e.g.,][]{OD_2011, Shetye_etal_2021, Bate_etal_2022}

\begin{figure*}[tp]
 \epsscale{0.9}
\plotone{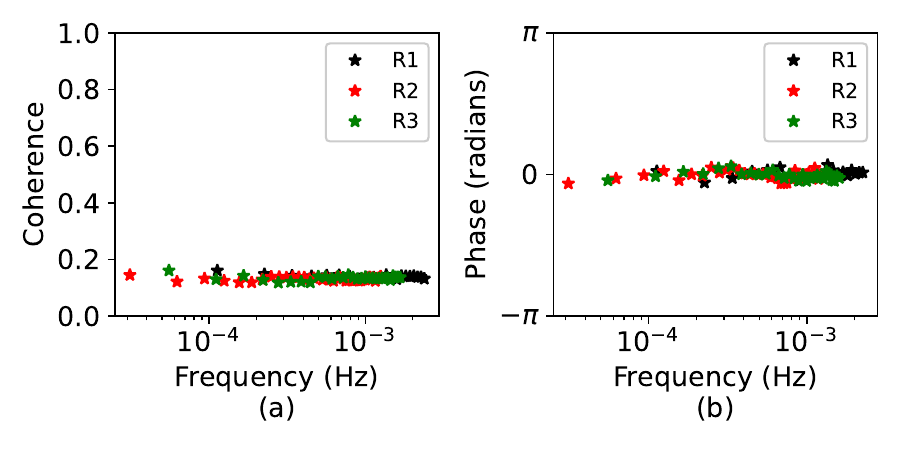}
    \caption{Cross-spectral analysis of Doppler velocity and intensity. Panel (a): Spatially averaged coherence versus frequency for the three selected regions, R${1}$, R${2}$, and R$_{3}$. Panel (b): Corresponding spatially averaged phase differences versus frequency for the same regions.}
  \label{fig:fig_Coh}
\end{figure*}

\subsubsection{Model of Power Spectrum}
\label{sec:fit}

To characterize the frequency distribution of Doppler velocity fluctuations, we fit a parametric model to the observed power spectra. The model consists of a power-law background plus a localized enhancement represented by a log-normal function, and includes a high-frequency noise floor.
The model takes the form:
\begin{equation}
    P_{\mathrm{M}}(f_{j})= A f_{j}^{\alpha}+B+C \exp\!\bigg(-\frac{(\log f_{j}-D)^2}{2E^2}\bigg),
\end{equation}

Here, $f_{j}$ are the Fourier frequencies, $\alpha$ sets the slope of the underlying power-law background, $A$ its amplitude, $B$ the high-frequency noise floor, and $(C,D,E)$ describe the location, width, and strength of the enhancement. 

Best-fit parameters were obtained using a maximum-likelihood approach, which is appropriate for exponentially distributed periodogram values \citep{Anderson_etal_1990}. The method, likelihood formalism, and uncertainty estimation follow \citet{Tajfirouze_etal_2025}. 

Table~\ref{tab:fitparam} summarizes the best-fit parameters for the IRIS Si IV 1393.75~\AA\ spectra, and the fitted models are overplotted on the observed spectra in Figure~\ref{fig:fig_Doppler}b. This approach enables direct comparison of the spectral characteristics of TR plasma at coronal heights with previous coronal observations of Alfv\'enic velocity fluctuations.

The index $\alpha$ characterizes the power-law component of the spectrum, which we interpret as being consistent with a background cascade of Alfv\'enic turbulence.
% which we interpret as arising from a background cascade of Alfv\'enic turbulence.
Deviations from a pure power law, represented by additional components in the model, may indicate the presence of coherent Alfv\'enic waves superposed on this turbulent background. The inferred power-law index may be affected by noise, particularly at high frequencies where instrumental effects become more significant. However, the model includes a constant noise floor, ensuring that the fitted spectral index of the power-law component remains robust.

\begin{figure*}[tp]
 \epsscale{0.9}
\plotone{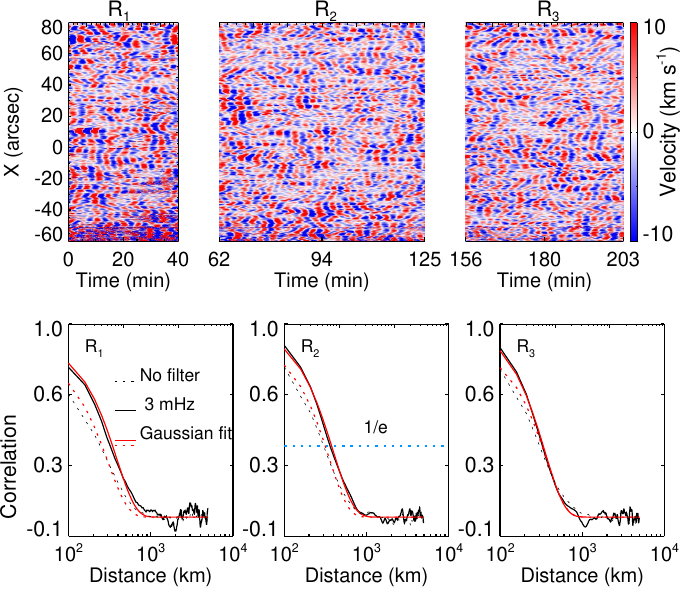}
    \caption{Upper panel: Filtered velocity time–distance diagrams for the three selected regions, $R_{1}$, $R_{2}$, and $R_{3}$.  
Lower panel: Average correlation as a function of distance for each region (solid curves), with overplotted Gaussian model fits (red curves). The average correlation computed over the full frequency spectrum (i.e., without filtering) is also shown (black dashed curve), together with its corresponding Gaussian fit (red dashed curve)}.
  \label{fig:fig_corr}
\end{figure*}

\subsection{Cross-Spectral Analysis}

The absence of any significant correlation between velocity and intensity (or line-width) fluctuations suggests that compressive wave modes do not contribute significantly to the measured signals \citep[e.g.,][]{Tajfirouze_etal_2025}, as such modes are expected to produce coupled velocity and density perturbations, and thus correlated intensity variations. This is consistent with expectations for transverse, Alfv\'enic–like motions, although similar power-spectral characteristics may also arise from other non-compressive processes \citep{DeMoortel_Pascoe_2012, Pant_etal_2019}.

To quantify the relationship between velocity and intensity fluctuations, we employ cross-spectral analysis \citep[see][for a description of the method]{Tajfirouze_etal_2025}. At each spatial location along the slit, we compute the magnitude-squared coherence and phase between the paired Doppler-velocity and intensity time series. The magnitude-squared coherence between two signals, $g$ and $h$, is defined as

\begin{equation}
    C^2_{gh}=\frac{\abs{P_{gh}}^2}{P_{gg}\,P_{hh}}
\end{equation}
where $P_{gh}$ is the cross-spectral density estimate of $g$ and $h$ and $P_{gg}$ and $P_{hh}$ are the power-spectral density estimates of the two signals. The phase difference can be estimated as

\begin{equation}
    \theta=\arctan\!\bigg(\frac{\mathrm{Im}(P_{gh})}{\mathrm{Re}(P_{gh})}\bigg)
\end{equation}
Here, the numerator and denominator correspond to the imaginary and the real parts of the cross-spectral density respectively. 

To reduce variance in the cross-spectral estimates, we apply the Welch method \citep{Welch_1967}, dividing each time series into overlapping segments and averaging the resulting spectra. 
The segment lengths were chosen so that each time interval contains approximately the same number of segments (five per interval), resulting in 80 exposures for region 1, 42 exposures for region 2, and 60 exposures for region 3. Overlapping segments ensure smooth spectral estimates, and values at the Nyquist frequency are excluded to avoid potential edge effects.

The coherence and phase results for all three time intervals are nearly identical (Figure~\ref{fig:fig_Coh}), demonstrating that the statistical properties of the fluctuations remain stable over the observation period. Coherence values are low, with a sharp peak near 0.13, and the spatially averaged phase differences are near zero. This combination indicates a lack of significant correlation between intensity and Doppler velocity, implying that the observed Doppler velocity fluctuations are predominantly incompressible, consistent with transverse Alfv\'enic motions reported in previous coronal and chromospheric observations \citep[e.g.,][]{Tajfirouze_etal_2025, Morton_etal_2025, DeMoortel_Pascoe_2012, Pant_etal_2019}. The quasi-stationarity of the signals further ensures that the time-series analysis yields robust and reproducible spectral estimates.

\subsection{Correlation along the Slit}\label{subsec:Correlation_length}

To quantify the spatial organization of the transverse velocity fluctuations, we estimate the perpendicular correlation length scale, $L_{\perp}$, which characterizes the transverse coherence of the Alfv\'enic motions and provides an observational constraint for models of wave-driven turbulence and energy transport \citep[e.g.,][]{CV_2005, VanB_2011}. In this context, the correlation length represents the characteristic transverse scale over which velocity fluctuations remain phase-coherent, and is therefore related to, but not necessarily identical with, the outer scale governing nonlinear interactions and energy transfer. The analysis follows the procedure described in \citet{Tajfirouze_etal_2025}; here we summarize only the aspects specific to the present IRIS observations.

Because IRIS sit-and-stare measurements provide a one-dimensional velocity field, spatial correlations are evaluated along the slit. At the observed polar limb heights, spicule-associated magnetic fields are expected to be approximately radial. Thus, the slit samples plasma primarily across the magnetic field direction in the plane of the sky. Under this viewing geometry, the derived correlation scale is interpreted as an estimate of the transverse coherence length of the motions, while any residual field-aligned component would lead to bias the measurement toward larger values of $L_{\perp}$.

To isolate the dominant wave band, Doppler velocity time series were filtered using a narrow Gaussian centered on the power enhancement near 4.5 mHz. The exact choice of central frequency within the enhancement band does not significantly affect the resulting correlation-length estimates. For each pixel, we computed the Pearson correlation between its filtered velocity time series and those of neighboring pixels along the slit. Averaging over all pixels yields a mean correlation function as a function of separation.  For comparison, the same analysis was also performed using the unfiltered velocity time series, i.e., over the full frequency spectrum.
We estimate $L_{\perp}$ by fitting the mean correlation profile with a Gaussian and taking the $e^{-1}$ decay distance:
\begin{equation}
    L_{\perp}=\sqrt{2}\,\sigma
\end{equation}
where $\sigma$ is the fitted width parameter.
For the filtered velocity signals, the resulting perpendicular correlation lengths are approximately 500 km, 505 km, and 410 km for regions R$_{1}$, R$_{2}$, and R$_{3}$, respectively ( assuming that an arcsecond is equivalent to 1$^{\arcsec} \approx$ 727 km). We verified that varying the filter center within the broader 3–7 mHz enhancement produces only modest changes in the inferred correlation lengths and does not alter the conclusions presented here. In contrast, the correlation lengths derived from the unfiltered data are systematically smaller in all three regions (330 km for R$_{1}$, 340 km for R$_{2}$, and 380 km for R$_{3}$). This indicates that the narrowband filtering isolates more spatially coherent velocity fluctuations, while the inclusion of the full frequency spectrum introduces additional, less correlated components that reduce the overall coherence scale. This behavior is consistent with the presence of multiple wave components and/or turbulent fluctuations contributing over a range of spatial and temporal scales.
The inferred coherence length, $L_{\perp} \approx 300$–$500$ km, is comparable to typical Type II spicule widths \citep[$\sim 200$ km; e.g.,][]{DePontieu_etal_b_2007}, This similarity suggests that the measured scale may reflect coherent transverse motions across individual spicule structures, such as kink-like oscillations, in addition to any role it may play as a characteristic scale in turbulence-based interpretations. Therefore, the measured $L_{\perp}$ should be regarded primarily as a coherence scale of the observed Doppler velocity fluctuations rather than a unique measure of a turbulence outer scale.
% suggesting that transverse Alfv\'enic motions remain coherent across the full cross-section of individual spicules and potentially over unresolved sub-structure within them.
\begin{figure}[tp]
 \epsscale{1.1}
\plotone{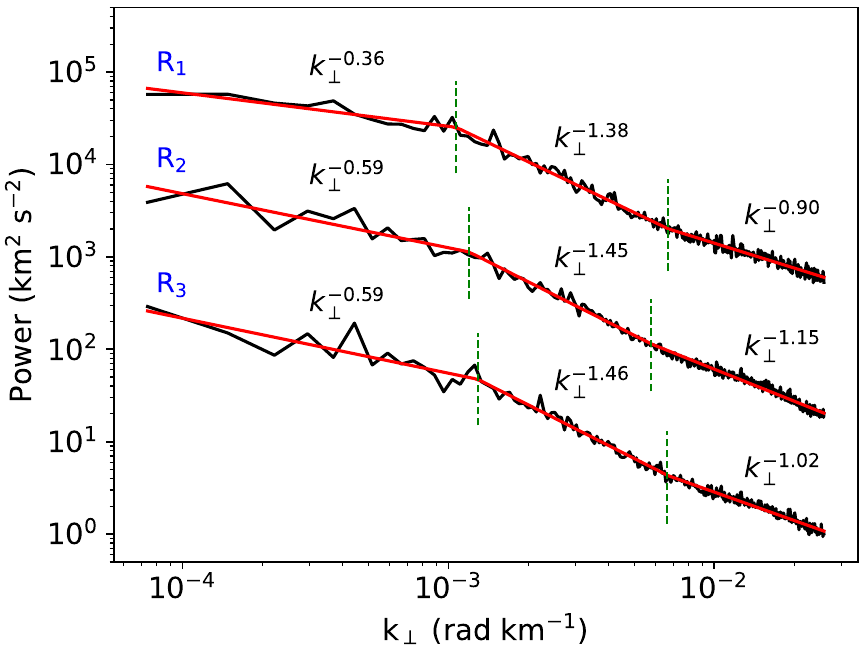}
    \caption{The spatial power spectra for three different time intervals marked in Figure~\ref{fig:fig_Doppler}. Top, middle, and bottom panels correspond to regions R1, R2, and R3, respectively. For better visualization, the R2 and R3 power spectra are vertically shifted by multiplicative factors of 0.06 and 0.003, respectively.} 
  \label{fig:fig_perp_spectra}
\end{figure}

\begin{figure*}[t]
  \centering
  \includegraphics[width=\textwidth]{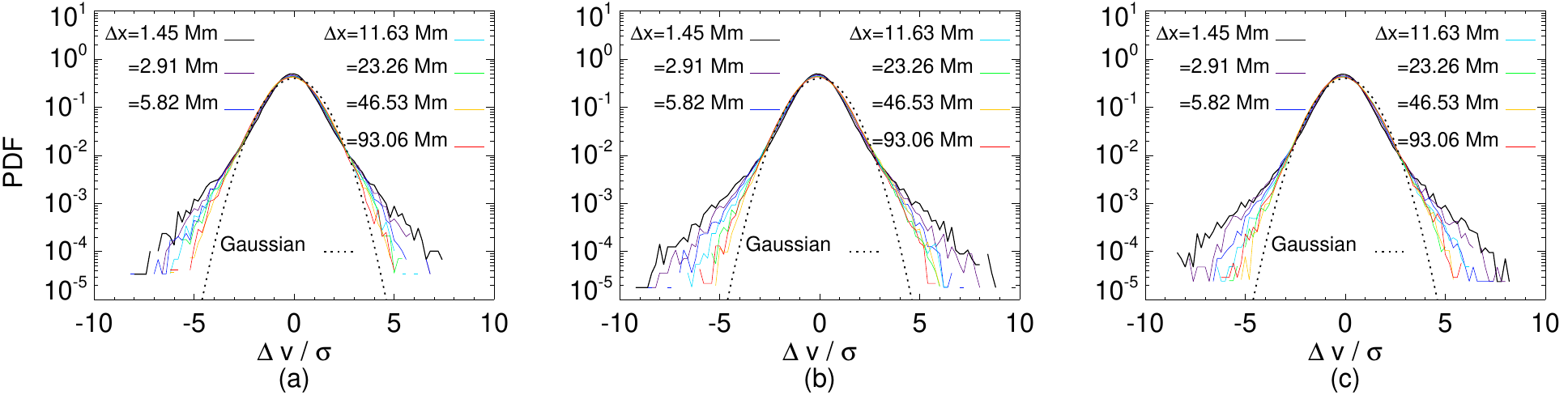}
  \caption{Probability density functions (PDFs) of the normalized increments for different spatial increments. Panels (a), (b), and (c) correspond to regions R1, R2, and R3, respectively, as marked in Figure~\ref{fig:fig_Doppler}. The colored solid lines show the PDFs at different spatial increments, while the dotted line represents a Gaussian distribution with unit rms for comparison. }
  \label{fig:fig_intermittency}
\end{figure*}

These measurements provide a direct observational constraint on the transverse coherence of Alfv\'enic fluctuations in spicule plasma extending into the low corona, highlighting the importance of explicitly distinguishing coherence scales from other characteristic length scales reported in previous studies. Our Doppler-velocity correlation lengths are smaller than the chromospheric intensity-based correlation lengths reported by \citet{Bailey_etal_2025}. This difference likely reflects the distinct physical quantities being measured, with intensity fluctuations tracing the coherence of emitting plasma structures and Doppler velocities tracing the coherence of wave-related motions.

\subsection{Spatial Analysis}
\subsubsection{Perpendicular Power Spectra and Intermittency of Velocity Increments}

To characterize the scale-dependent distribution of energy in the low corona, we computed the spatial power spectra using a Fourier transform of the velocity field. To mitigate spectral leakage arising from non-periodic boundary conditions, the data were apodized using a Tukey window function. We adopt a mild taper ($\alpha=0.1$), which minimizes edge-induced distortions while preserving the intrinsic large-scale power distribution. The spectra were normalized by the mean squared window power to ensure correct amplitude scaling. We verified that the inferred spectral slopes and break locations are only weakly sensitive to the precise choice of window parameter within reasonable values. Figure~\ref{fig:fig_perp_spectra} shows the spatial power spectra for three separated regions marked in Figure~\ref{fig:fig_Doppler}(a). For visualization, the second and third spectra have been shifted vertically. Each spectrum exhibits distinct spectral breakpoints that can be interpreted, in the context of turbulence, as transitions between different physical regimes. Based on these breakpoints, and adopting a turbulence framework for interpretation, we identify three ranges: a large-scale range associated with energy injection, an intermediate range characterized by scale-invariant energy transfer, and a small-scale range where the spectrum flattens. The extent to which this small-scale range is affected by instrumental noise is not currently clear, therefore we do not interpret this physically. The spatial power spectra were fitted using a broken power-law model with three segments. The breakpoints and slopes were treated as free parameters and determined using nonlinear least-squares fitting. The model is constructed to be continuous between segments, ensuring a consistent representation of the spectrum. This approach allows the break scales and spectral slopes to be estimated directly from the data.

The slopes of the spectra are marked in Figure~\ref{fig:fig_perp_spectra} and the average inertial range value is $-1.43$, slightly shallower than both of the commonly predicted values, $-5/3$ and $-3/2$, for homogeneous strong MHD turbulence \citep{Schekochihin_2022} but consistent with the range of values seen in simulations of reflection-driven turbulence in the solar corona \citep{P-Ch_2013, CH-P_2019}.

In turbulent systems, velocity increments are expected to be approximately Gaussian at large spatial scales, while developing increasingly non-Gaussian, heavy-tailed distributions at smaller scales due to intermittency and coherent structures.
To probe the signature of turbulent dynamics, we analyzed the intermittency of velocity increments, $\Delta v=v(x+\Delta x)-v(x)$, for the selected regions by examining their probability distribution functions (PDFs) at varying spatial lags, $\Delta x$. The spatial increments were chosen to increase on a logarithmic scale. Figure~\ref{fig:fig_intermittency} shows the PDFs of velocity increments normalized by their root mean square (rms) values. The PDFs exhibit clear non-Gaussian features at small spatial scales, characterized by heavy tails, consistent with intermittent turbulence. As the spatial lag increases, the PDFs gradually transition toward Gaussian statistics, reflecting the cumulative effect of multiple independent wave packets.

\begin{figure*}[t]
  \centering
  \includegraphics[width=\textwidth]{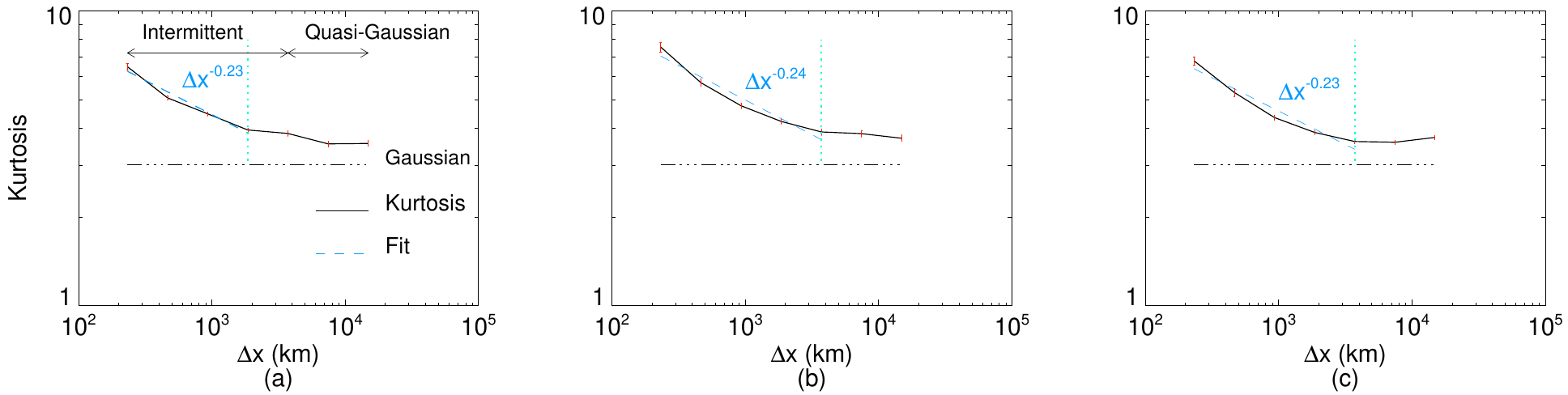}
  \caption{The kurtosis as a function of spatial increments. Panels (a), (b), and (c) correspond to regions R1, R2, and R3, respectively, as marked in Figure~\ref{fig:fig_Doppler}. The solid lines show the kurtosis. The red lines indicate the associated uncertainties. The green vertical dashed lines separate the quasi-Gaussian regime and the intermittent regime. The blue dashed lines show a power-law fit to the intermittent regime.}
  \label{fig:fig_kurtosis}
\end{figure*}

We computed the kurtosis of velocity increments as a function of spatial increment to quantify the deviations from Gaussian statistics and to characterize the nature of intermittency. Kurtosis is defined as
\begin{equation}
    K(\Delta x)=\frac{\braket{(\Delta v)^4}}{\braket{(\Delta v)^2}^2}
\end{equation}
where $\Delta v$ is the velocity increment. For a Gaussian distribution, $K = 3$, while deviations from this value indicate the presence of intermittency and non-Gaussian statistics.
As shown in Figure~\ref{fig:fig_kurtosis}, kurtosis systematically decreases with increasing spatial increments, indicating stronger intermittency and more pronounced bursty fluctuations at smaller scales, which gradually weakens toward larger scales. The breakpoints in the kurtosis profiles mark a transition from quasi-Gaussian to intermittent statistics. These characteristic scales are comparable, but not identical, to the correlation lengths derived from the cross-correlation analysis, which quantify the spatial coherence of fluctuations. The values inferred from the kurtosis are approximately 1940 km for R$_1$, and 3880 km for R$_2$ and R$_3$, and are larger than those obtained in Section~\ref{subsec:Correlation_length} by about an order of magnitude. This discrepancy may reflect the fact that while these diagnostics all provide an order of magnitude estimate of the outer scale, they probe slightly different aspects of the fluctuations, as well as being influenced by finite data length and noise, which can affect correlation-based estimates \citep{Perez-AGU_2013, Isaacs_etal_2015}.

We fitted a power law to the intermittent part of the kurtosis plots. It can be seen in Figure~\ref{fig:fig_kurtosis} that the slope is $\sim$ -0.23 for all three regions, which is close to the values found in turbulence analysis of solar wind data \citep[see e.g., ][]{Mondal_etal_2025}, supporting an interpretation that the background power law spectrum is consistent with Alfv\'enic turbulence.

These scale-dependent non-Gaussian statistics, combined with the scale-dependent power-law behavior observed in the spatial and temporal power spectra, are consistent with a picture of intermittent, spatially coherent Alfv\'enic turbulence in tall polar spicules. The decreasing kurtosis with scale further implies that small-scale structures and localized bursts play a significant role in the dynamics and energy dissipation processes consistent with expectations for intermittency and scale-dependent energy transfer in Alfv\'enic turbulence.

The spatial coherence scales derived from cross-correlation analysis, the spectral breakpoints in the perpendicular power spectra, and the transition scales identified from the kurtosis profiles are found to differ systematically. In particular, the correlation length is smaller than the characteristic scales associated with the spectral and intermittency diagnostics. This reflects the fact that the different methods probe distinct statistical properties of the velocity field, including phase coherence, energy distribution across scales, and the emergence of intermittent structures.

\medskip

\section{Discussion and Conclusion} \label{sec:discussion}

Our analysis of transverse velocity fluctuations in polar spicules using IRIS sit-and-stare observations provides a detailed characterization of the spatial and temporal properties of Alfv\'enic fluctuations in the transition region and low corona. Previous studies have extensively quantified wave amplitudes, energy flux, and propagation properties of Alfv\'enic motions \citep[e.g.,][]{DePontieu_etal._2007, Jess_etal_2009, Tavabi_etal_2015, Morton_etal_2015, Morton_etal_2025, Tajfirouze_etal_2025}. In this work, we extend this framework by combining spatial coherence, spectral analysis, and higher-order statistics to investigate the multi-scale structure of these fluctuations in spicules.
We constrain their wave power distribution, spatial coherence, and intermittency.

We find that the velocity fluctuations are predominantly incompressible and show no significant correlation with intensity variations, consistent with earlier observational and theoretical expectations for Alfv\'enic motions in the corona. A cross-correlation analysis indicates that the fluctuations remain coherent over transverse spatial scales of $L_{\perp} \approx 300$–$500$ km, comparable to typical Type II spicule widths, suggesting coherence across the full spicule cross-section. These values are approximately an order of magnitude smaller than the characteristic scales inferred from the break points in the spatial power spectrum and kurtosis analysis of the velocity field. This difference is expected because the three diagnostics quantify different properties of the velocity field. The correlation length measures the decay of similarity between fluctuations at different positions and is therefore primarily sensitive to phase coherence of the dominant velocity fluctuations. 
In addition, the measured Doppler-velocity correlation lengths are comparable to typical spicule widths. This suggests that the observed coherence scale may partly reflect the collective nature of kink-like transverse motions within individual magnetic structures. Consequently, the measured correlation length should be interpreted primarily as a coherence scale of the observed fluctuations and not uniquely as a turbulence outer scale. In particular, coherent kink-like motions within individual spicules could naturally produce transverse correlation lengths comparable to the observed values.

In contrast, the spectral break reflects changes in the distribution of energy across spatial scales, while the kurtosis transition is controlled by the emergence of intermittent, high-amplitude structures that dominate higher-order statistics. These structures can persist over larger spatial separations than the phase coherence scale, leading naturally to larger characteristic lengths from the spectral and intermittency-based diagnostics. The resulting separation of scales is therefore consistent with expectations for an inhomogeneous, intermittent turbulent system, in which coherence, energy transfer, and intermittency are not characterized by a single unique spatial scale.

The temporal power spectra exhibit broadband power-law behavior extending to $\sim$10 mHz, {together with a broad enhancement spanning approximately 3–7 mHz and peaking near 4–6 mHz}. This enhancement overlaps with the global 5-minute oscillation band observed across multiple layers of the solar atmosphere \citep{DePontieu_2005, Didkovsky_etal_2011, Morton_etal_2025A, Guglielmi_etal_2015, Huang_etal_2025}. Using IRIS observations of tall polar spicules, we probe chromospheric and transition-region plasma at low coronal heights, where enhanced power in the five-minute band remains evident in the Doppler velocity fluctuations. This suggests that oscillatory power originating in the lower atmosphere can persist to coronal heights within spicular structures \citep[see also,][]{Qi_etal_2026}.

Spatial Fourier spectra of the Doppler velocity fluctuations exhibit a clear power-law interval $\sim k_\perp^{-1.43}$. This slope is slightly shallower than the classical expectations of $-5/3$ and $-3/2$ associated with homogeneous strong MHD turbulence, but is consistent with the range of values seen in simulations of reflection-driven turbulence in the corona. While reflection-driven processes may contribute to this behavior, the extent to which coherent chromospheric features (e.g., spicules or wave-driven structures near $\sim$4 mHz) influence the measured $k_\perp$ spectrum remains unclear. A more precise characterization of the perpendicular turbulent cascade, and its separation from structured or wave-dominated signals, is required to establish the physical origin of the observed spectral slope.

Higher-order statistics further support this picture. Velocity increment probability distribution functions show clear non-Gaussian behavior, with enhanced kurtosis at small spatial scales that decreases toward larger scales, indicating a transition from intermittent, bursty fluctuations to more Gaussian-like behavior. Such scale-dependent intermittency is commonly associated with turbulent cascades and is here identified observationally in spicule dynamics, providing further support for turbulence-based interpretations of coronal wave evolution. Together with the measured spectral slopes and finite coherence lengths, this behavior is consistent with nonlinear interactions that may transfer energy to smaller spatial scales through a turbulent cascade.

The combined diagnostics from coherence, spectral structure, and intermittency indicate that the observed fluctuations are consistent with organized Alfv\'enic dynamics rather than random or purely projection-induced signals. While the fluctuations are largely incompressible, contributions from compressible perturbations cannot be fully excluded due to line-of-sight integration and finite spatial resolution; however, these effects are not expected to significantly alter the inferred properties. This strengthens the interpretation that the measured signals predominantly reflect transverse Alfv\'enic dynamics rather than observational artifacts.

While it is possible that other processes may contribute to the results measured here, the combination of Alfv\'enic motions, broadband temporal spectra, spatial power-law behavior, and non-Gaussian velocity increment statistics indicates that most likely a cascade of Alfv\'enic turbulence is present.

Taken together, the results indicate that transverse motions in polar spicules exhibit coherence over megameter scales while also showing structured variability at smaller scales. Overall, this work provides new multi-diagnostic observational constraints on the spatial structure, intermittency, and spectral properties of Alfv\'enic fluctuations in the low corona. These results have direct implications for models of wave-driven turbulence, particularly those incorporating reflection, structuring, and inhomogeneity, and contribute to a more complete picture of how wave energy is transported and redistributed in the solar atmosphere.

\medskip

\begin{acknowledgements}
E.T. and C.H.K.C. were supported by UKRI Future Leaders Fellowship MR/W007657/1.
C.H.K.C. was also supported by STFC Consolidated Grant ST/X000974/1.
R.J.M. would like to thank the UKRI for financial support via a UKRI Future Leader Fellowship
(RiPSAW MR/T019891/1). P.R.Y. acknowledges support from the GSFC Internal Scientist Funding Model competitive work package program and the Hinode project. E.T. would like to thank P. Antolin, B. De Pontieu, N. Dadashi, D. Manzini, C. Waters, and S. Mondal for their valuable discussions.
\medskip 

\textit{Data Availability:} IRIS data are publicly available at \url{https://iris.lmsal.com/search/}.

\medskip

\end{acknowledgements}

\medskip 
%\clearpage

% \bibliographystyle{aasjournal}
% \bibliography{bibtex}

\end{document}